# Review:
# Towards Spintronic Quantum Technologies with Dopants in Silicon

Gavin W. Morley

Department of Physics, University of Warwick, Coventry CV4 7AL,
gavin.morley@warwick.ac.uk

## Abstract

Dopants in crystalline silicon such as phosphorus (Si:P) have electronic and nuclear spins with exceptionally long coherence times making them promising platforms for quantum computing and quantum sensing. The demonstration of single-spin single-shot readout brings these ideas closer to implementation. Progress in fabricating atomic-scale Si:P structures with scanning tunnelling microscopes offers a powerful route to scale up this work, taking advantage of techniques developed by the computing industry. The experimental and theoretical sides of this emerging quantum technology are reviewed with a focus on the period from 2009 to mid-2014.

## Table of Contents

**X.1 Introduction to Phosphorus Dopants in Silicon (Si:P) as a Model System for Electron Paramagnetic Resonance (EPR)**

Phosphorous, arsenic, antimony and bismuth are in group V of the periodic table, and they can enter the silicon lattice substitutionally, as a nucleus with a positive charge and a spare electron. At room temperature this spare electron is donated to the conduction band, so these dopants are referred to as donors. The phosphorous donor is of great technological importance as it is used to negatively dope silicon in modern electronics. At helium temperatures the positive nuclear charge of these donors captures a spare electron providing an analogue for the hydrogen atom.

The electron paramagnetic resonance of phosphorus and arsenic dopants in silicon was observed in Bell Labs in the 1950s[1]. Bell Labs' George Feher then studied these materials intensely, revealing exchange-coupled clusters[2], long electron spin-lattice relaxation times[3] ($T_1 > 1000$s) which were affected by applied stress[4], and the electronic structure[5] using his electron nuclear double resonance (ENDOR)[6] technique. At the same time, also in Bell Labs, Gordon and Bowers observed the first microwave electron spin echoes, using phosphorus and lithium dopants in silicon[7]. The electron $T_2$ time they reported in 1958 for $^{28}$Si:P was 0.5 ms, prompting the authors to suggest that this could be a good system for storing information, following an earlier proposal[8]. Kane extended this idea in 1998 by proposing a $^{28}$Si:P quantum computer[9].

This chapter reviews developments in donor quantum information research between 2009 and 2014. Some previous reviews of silicon qubits have considered both donors and gate-defined quantum dots[10-11] and silicon qubits have also been reviewed in the context of other qubit implementations[12-14].

**X.2 Requirements and Proposals for Quantum Technologies**

The goal of building a quantum computer has always driven quantum technology research, but other applications such as quantum-enhanced sensing of magnetic fields, electric fields and temperature are now attracting increasing interest. Quantum metrology uses techniques such as entanglement to achieve higher precision measurement than the classical shot-noise limit[15]. Quantum technologies generally require[16] that the quantum system can be (1) initialized into a useful starting state and (2) controlled with high fidelity faster than (3) the timescale for loss of quantum coherence ($T_2$). The fourth requirement is a readout of the classical state of the system (is the spin up or down?) on a timescale faster than the loss of this classical information ($T_1$).

**X.2.1 Qubit Initialization (Spin Polarization)**

The donor electron spin and the nuclear spins coupled to it, whether from the donor nucleus or nearby $^{29}$Si, could serve as qubits. Prior to running a quantum computation (or sensing something in the environment) these spins should be initialized to a useful starting state, such as with all spins polarized. An impressive route to this using bound exciton transitions is described below in Section X.4, but a simple alternative is to apply a large magnetic field and a low temperature[17]. This equilibrium polarization has been used to reach over 99.9% electronic polarization[18] at temperatures of 1.4 K, and some of this has been transferred to the $^{31}$P nuclear spin, providing up to 68% nuclear polarization[18-19]. High fields up to 12 T have been used with EPR frequencies of up to 336 GHz [17, 19], and this has been combined with electrically detected EPR experiments to study devices with high sensitivity[17-22]. Si:P in very high magnetic fields (exceeding 30 T) has been studied with far-infrared spectroscopy at 2.2 K, providing an analogy for hydrogen on white dwarf stars[23].

A thermal electronic polarization of over 99.9999999% is expected by using a temperature of 100 mK and a magnetic field of 1.6 T, and this approach has been integrated with single-spin single-shot readout using a single electron transistor[24-27] as described below in Section X.2.4. This readout itself can also be used to polarize qubits, as any qubits found to be in the wrong state can be re-measured until they are in the desired state.

**X.2.2 Spin Qubit Control**

Single qubit control for electronic and nuclear spins is achieved with magnetic resonance pulses[28]. For many quantum technology schemes it is desirable to be able to selectively address particular qubits by their frequency, which could be achieved in several ways, including by making use of multiple donors species as they have different hyperfine couplings and nuclear spin. It would be more useful for scaling up to many qubits if this selectivity could be controlled by a gate during a quantum computation. Kane envisioned doing this by applying an electric field and using the Stark effect[9]; while this effect has proved to be weak for the group V donors studied to date, lithium donors are predicted to have a large Stark effect [29]. Applying strain currently seems to be the most promising way to selectively address Si:P using a gate as the EPR resonance has been shifted by more than the $^{28}$Si:P resonance width[30]. Piezo-electric actuators could apply this strain locally as shown in figure X.1.

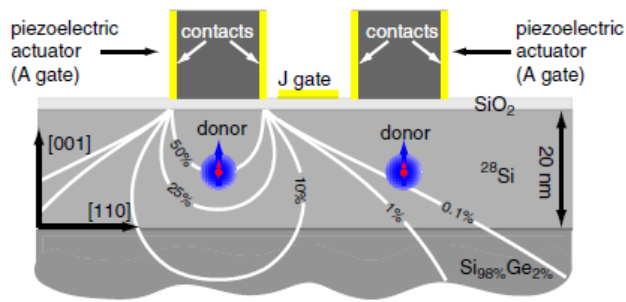

*Figure X.1. Schematic of a proposal for strain control of donor qubits. Reprinted[30] figure with permission from L. Dreher, T. Hilker, A. Brandlmaier, S. Goennenwein, H. Huebl, M. Stutzmann and M. Brandt, Physical Review Letters, 106, 037601 (2011). Copyright 2011 by the American Physical Society.*

In one set of experiments, two single-phosphorus-atom devices were found to have very different hyperfine couplings of 116.6 MHz and 96.9 MHz, and the difference between these values is over 10,000 times bigger than the linewidth of the spin resonance transitions, suggesting superb opportunities for spectral selectivity[27].

Coupling two donor atoms while retaining their long coherence times remains a challenge. A strongly interacting pair of arsenic donors has been studied where an exchange coupling of around 180 GHz was inferred[31]. A pair of exchange-coupled phosphorous dopants have been studied[32], but it was found that their strong coupling of ~70 GHz led to a relatively fast triplet-to-singlet relaxation time of $T_1$ ~ 4 ms. Using a coupling between donors that is less than the hyperfine coupling ($A$ ~ 117 MHz for Si:P) should bring back the long $T_1$ times that are wanted for quantum computation[33]. Alternative proposals for coupling donors in silicon include the use of a ferromagnet[34], phonons[35], photons[36] and the Rydberg excited state of donor electrons[37]. Coherent excitation of the Rydberg states of Si:P has been demonstrated but the coherence time is only 160 ps [38]. Entanglement between the hyperfine-coupled electronic and nuclear spins of a Si:P ensemble has been demonstrated[39] with ENDOR at 3.4 T.

### X.2.3 Spin Qubit Coherence Times

The long spin coherence times[7] seen in the 1950s were a strong motivation to study qubits in silicon[9], but recent progress has extended these times by orders of magnitude. One key step has been to reduce the density of spins in the silicon crystal: further reductions in $^{29}$Si have been beneficial, as well as lower donor densities, as shown in Figure X.2[40]. Measurements in Princeton University of a $^{28}$Si:P ensemble (for a crystal with <50ppm residual $^{29}$Si and donor density $10^{14}$ cm$^{-3}$) produced an electron spin $T_2$ of 10 seconds at 1.8 K, by extrapolating spin echo decay measurements to the limit where the refocusing pulse had zero length to remove the effects of instantaneous diffusion[41]. It was still necessary to use magnitude detection (collecting only the amplitude, rather than the in-phase and quadrature components of the echo) to suppress phase noise in the EPR spectrometer.

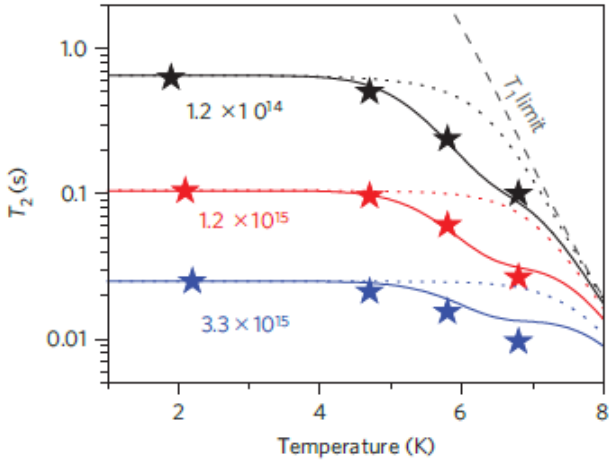

*Figure X.2. Electron spin $T_2$ times for $^{28}$Si crystals with three different donor densities after extrapolating to the limit of zero duration refocusing pulse to remove instantaneous diffusion. The coherence time of the lowest concentration sample was further extended by applying an external magnetic field gradient to suppress donor flip-flops. Reprinted[41] by permission from Macmillan Publishers Ltd: A. M. Tyryshkin et al., Nature Materials **11**, 143, copyright 2012.*

Nuclear spins have a smaller magnetic moment than electrons so are generally found to have longer relaxation times. The most recent report of nuclear coherence times in $^{28}$Si was $T_2 = 3$ hours at helium temperatures (as shown in Figure X.3), and 39 minutes at room temperature[42] for ionized phosphorous donors, making use of dynamic decoupling[43]. These experiments used low dopant densities (only ~$5 \times 10^{11}$ phosphorous cm$^{-3}$) which were detectable because of the Auger electron detected magnetic resonance[44] which is described further in Section X.4 below on bound exciton transitions.

Long coherence times have also been measured for single donor spins in silicon, with single-shot readout, as discussed in the next section, X.2.4.

Simulations of a central spin in baths of $^{29}$Si and donors reproduce the shape and magnitude of the electron spin coherence decay[40, 45] with no fitting parameters. The cluster-correlation expansion[46] provides an efficient way to handle the large number of spins (over 1000), making use of the insight that decoherence from the bath is dominated by small clusters of bath spins.

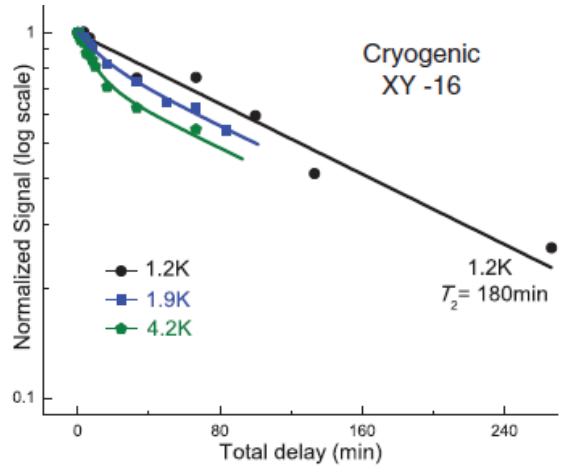

*Figure X.3. Si:P nuclear spin decoherence with XY-16 dynamic decoupling. The 1.9 K and 4.2 K data were fit using biexponentials, with the longer component set to 180 min. Reprinted[42] from K. Saeedi et al., Science, 2013, **342**, 830. Reprinted with permission from AAAS.*

**X.2.4 Single Spin Readout with Single Electron Transistors (SETs)**
Single-spin single-shot readout of an electron in silicon was achieved with a specially-fabricated single electron transistor (SET) at the University of New South Wales[24, 47]. The SET is shown in Fig. X.4, and is operated as a sensitive detector of electric charge, by voltage biasing it to Coulomb blockade where current cannot flow through the transistor. A charge moving in the environment of this SET changes the bias allowing current to flow from source to drain: this is the detected signal. Charge moves in this way when it tunnels from a single phosphorous donor to the SET island, and this process is spin-dependent in a magnetic field because if the phosphorous electron is in the higher-energy Zeeman state it has a greater chance of tunneling away.

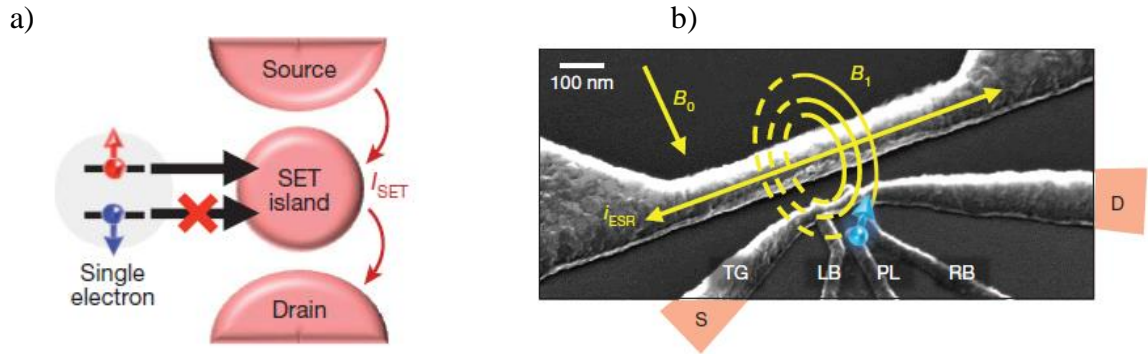

*Fig. X.4. a) Schematic of the single electron transistor (SET) used to readout a single electron spin in silicon[24]. Reprinted by permission from Macmillan Publishers Ltd: A. Morello et al., Nature, **467**, 687, copyright 2010 b) Scanning electron micrograph of a device similar to the one used in the experiment[25]. The SET (lower right portion) consists of a top gate (TG), plunger gate (PL), left and right barrier gates (LB and RB) and source/drain contacts (S and D). The microwave transmission line is shown in the upper left portion. The donor (blue dot) is subject to an oscillating magnetic field $B_1$ from the transmission line which is perpendicular to the in-plane external field $B_0$. Reprinted by permission from Macmillan Publishers Ltd: J. J. Pla et al., Nature, **489**, 541, copyright 2012.*

The SET readout permitted pulsed EPR[25] and then pulsed NMR[26] measurements on a single phosphorous dopant. This readout technology removed a key blockage in the silicon qubit field, and revealed encouraging spin coherence times for the electron and nuclear spins. The proximity of the donor to an oxide interface and nearby electrostatic gates did not introduce additional decoherence. Some coherence measurements are shown in Figure X.5, and the same group have even more recently reported[27] electron $T_2$ times of around 1 ms (with a spin echo) and 0.56 s (with dynamic decoupling), as well as a $T_2$ time for the $^{31}$P nucleus with a neutral donor of 1.5 ms for one device and 20 ms for another (both with a spin echo). Ionizing the donor provided a nuclear $T_2$ time of 1.75 s (spin echo) and 35.6 s (with dynamic decoupling). These times are shorter than those measured in bulk $^{28}$Si samples for electrons[41] and nuclei[44], which was attributed to Johnson-Nyquist thermal noise due to the microwave source[27]. High fidelity control pulses were achieved, reaching 97% for the electron and 99.99% for the nuclear spin.

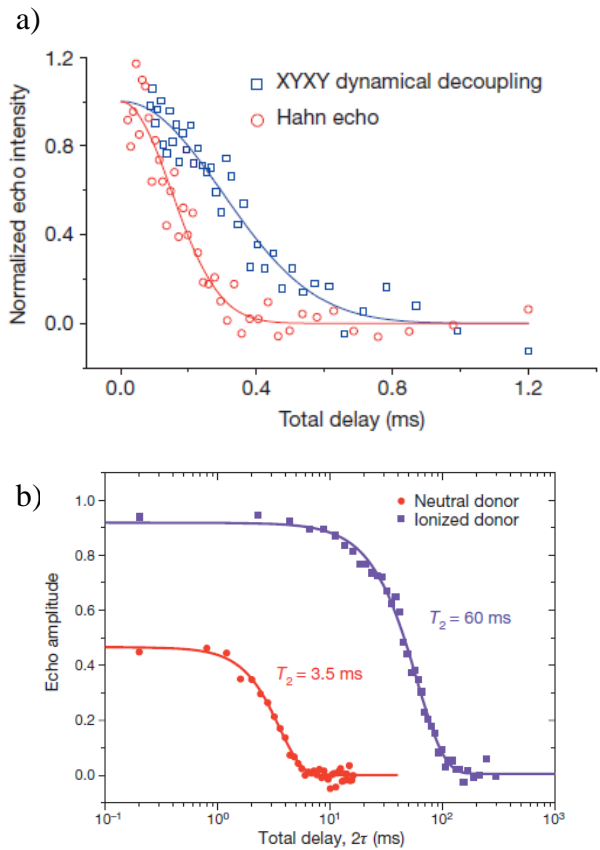

*Figure X.5. Coherence times of a $^{31}$P donor: a) electron $T_2$ time. Reprinted by permission from Macmillan Publishers Ltd: J. J. Pla et al., Nature, **489**, 541, copyright 2012. b) nuclear $T_2$ time. Reprinted by permission from Macmillan Publishers Ltd: J. J. Pla et al., Nature, **496**, 334, copyright 2013.*

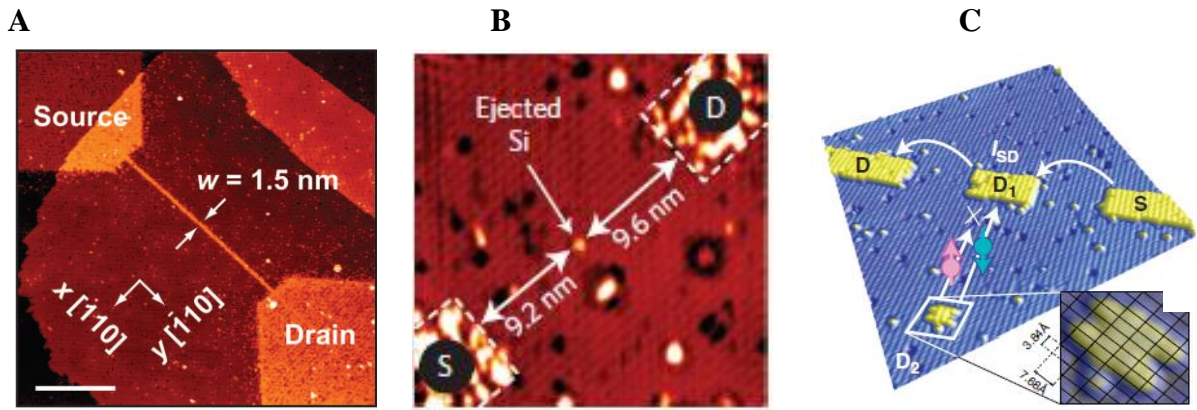

*Fig. X.6. Scanning tunneling microscope images of precision donor devices made with hydrogen lithography: A) A metal wire which displayed Ohmic conductivity[53]. From B. Weber et al., Science, 2012, **335**, 64. Reprinted with permission from AAAS. B) a single-atom transistor[54]. Reprinted by permission from Macmillan Publishers Ltd: M. Fuechsle et al., Nature Nanotechnology, **7**, 242, copyright 2012 and C) an SET to readout the state of the cluster of phosphorous atoms[55]. Reprinted by permission from Macmillan Publishers Ltd: H. Buch et al., Nature Communications, **4**, 2017, copyright 2013.*

### X.3 Atomic Scale Fabrication with Scanning Tunneling Microscopy (STM)

Phosphorous dopants can be placed into the silicon lattice with atomic precision using hydrogen lithography, using techniques developed at the University of New South Wales[48-49]. A scanning tunneling microscope (STM) is used to image the silicon surface, which is terminated with a layer of hydrogen atoms. The STM tip can be used to selectively remove hydrogen atoms, leaving gaps where phosphine molecules can dock, followed by surface chemistry to leave just the phosphorous atom behind[50]. After encapsulation in epitaxial silicon[51] these atomic precision devices can be contacted electrically[52]. This work has recently led to the fabrication of Ohmic metal wires made of a chain of phosphorous atoms[53], permitting the creation of a transistor whose current is controlled by a single phosphorous atom[54] as well as single-shot SET measurements of a cluster of phosphorus spin qubits[55] (see figure X.6). Most recently the group have measured exchange coupling in double donor systems[56] demonstrating a way to build and scale up the precise architectures required to achieve larger scale quantum information processing using donors in silicon.

### X.4 Bound Excitons for Dynamic Nuclear Polarization (DNP) and Spin Readout

Silicon does not have a direct band gap which generally precludes the kind of coherent optical experiments that are so useful for spin polarization and readout of nitrogen-vacancy spin qubits in diamond[57-58]. However, pioneering work at Simon Fraser University showed that optical photons could be used to polarize and readout the spin state of qubits in silicon as long as bound exciton transitions are used[59] (see Figure X.7). The optical light creates an electron-hole pair (exciton) which remains bound to the $^{31}$P. When isotopically pure $^{28}$Si:P is used, these transitions become sharp enough to resolve the nuclear spin state (see Figure X.7 B), and frequency-selective excitation then permits polarization of the electronic and nuclear spins[60]. The electron polarization reaches 97% and the nuclear polarization reaches 90% at 4.2 K with a magnetic field of just 84.5 mT as shown in figure X.8[44].

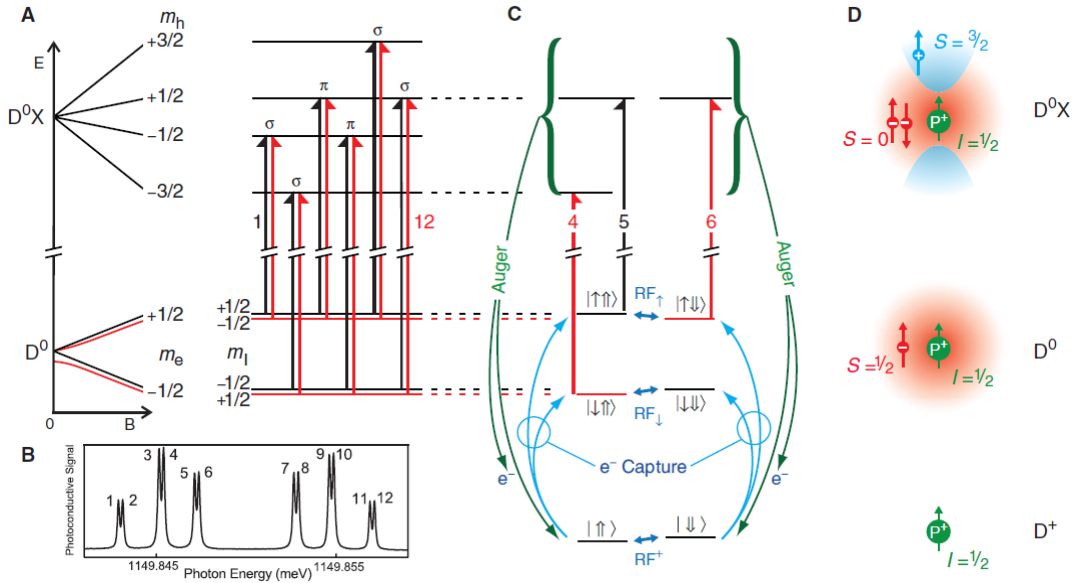

*Figure X.7. Energy levels and transitions of the P neutral donor ($D^0$), donor bound exciton ($D^0X$), and ionized donor ($D^+$)[42]. A) The Zeeman splittings of the $D^0$ and $D^0X$ states are shown from magnetic field B = 0 to B = 84.53 mT, along with the dipole-allowed optical transitions. B) Photoconductive readout spectrum without any $D^0$ hyperpolarization. C) The specific optical transitions (lines 4, 5, and 6) and nuclear magnetic resonance transitions ($RF_↑$, $RF_↓$, and $RF^+$) used to hyperpolarize, manipulate, and read out the nuclear spins. D) Sketches of the spins and charge densities of $D^+$, $D^0$, and $D^0X$. From K. Saeedi et al., Science, 2013, **342**, 830. Reprinted with permission from AAAS.*

After excitation of a bound exciton transition, the system decays back to the ground state with the ejection of an Auger electron, and non-contact electrical detection of these conduction electrons allows sensitive NMR experiments on bulk samples with a very low density of phosphorous qubits[44, 61]. This results in the longest coherence times of three hours (with ionization and dynamic decoupling at 1.2 K) for the $^{31}P$ nuclear spin[42] with a phosphorous concentration of just ~$5 \times 10^{11}$ cm$^{-3}$.

The spins of bismuth dopants in silicon (Si:Bi) have also been polarized with bound exciton spectroscopy[62-63].

In separate optical experiments without bound excitons, photoexcited triplet states of the oxygen-vacancy centre in silicon have been stored in $^{29}Si$ nuclear spins, providing access to nearly 100% spin polarization[64].

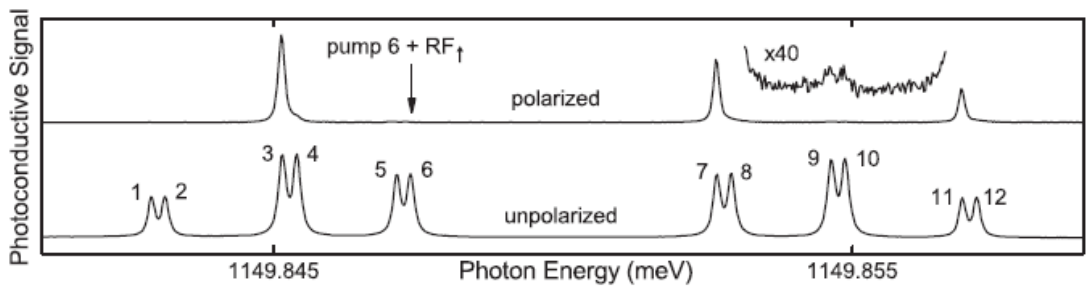

*Figure X.8. Photoconductivity spectra at T = 4.2 K and B = 84.5 mT, for the largely unpolarized equilibrium case (bottom) and using the hyperpolarization scheme (top).
The relative intensities of lines 3, 4, 9, and 10 give directly the relative populations of the $D^0$ electron and nuclear spin states[44]. From M. Steger et al., Science, 2012, **336**, 1280. Reprinted with permission from AAAS.*

## X.5 Bismuth Dopants in Silicon (Si:Bi)

Although some basic experiments were performed in the 1950s and 1960s (such as references 5 and 65), bismuth dopants in silicon have been much less studied than Si:P because of the key role of the latter in modern computer chips. The substitutional bismuth dopant has an electron spin of $S = ½$, like phosphorous, but the nuclear spin is $I = 9/2$ instead of $½$ and the hyperfine coupling is an order of magnitude larger at $A = 1.4754$ GHz. The large nuclear spin means that there is a larger Hilbert space in which quantum information can be stored.

In regimes where the Zeeman energy is comparable to the hyperfine coupling, specific magnetic field values (termed cancellation resonances in [66]) were shown to be of particular interest. These are close to optimal working points (OWPs) [66-68] where decoherence was found to be sharply suppressed[68]. This suppression of decoherence at OWPs has been observed experimentally[69] in both natural and enriched samples of silicon. OWPs can be close to or even coincident with "clock transitions", points where there is first order insensitivity to magnetic field values, but for $^{29}$Si spectral diffusion they do not exactly coincide[68]. Figure X.9 shows the Breit-Rabi energy levels in this region, and Figure X.10 shows this landscape of magnetic resonance transitions as a function of magnetic field and excitation energy. The same electron-nuclear mixing that lies behind these features has been shown to allow faster quantum control of the electron-nuclear system[70] in comparison with the standard unmixed regime used in Si:P (ref 71).

Electron spin echo envelope modulation (ESEEM)[72] and ENDOR[68] have been used to study the overlap of the donor wavefunction with naturally-occurring $^{29}$Si. Low-field (6-110 mT) measurements with 50 MHz and 200 MHz excitation showed that the bismuth excitation energy could be tuned for future coupling with superconducting qubits[36, 73], which require low magnetic fields. Si:Bi has electron spin coherence times that are at least as long as Si:P with natural silicon[74-75] and isotopically pure $^{28}$Si (ref 76). Like Si:P, Si:Bi is suitable for bound exciton experiments including nuclear hyperpolarization[62].

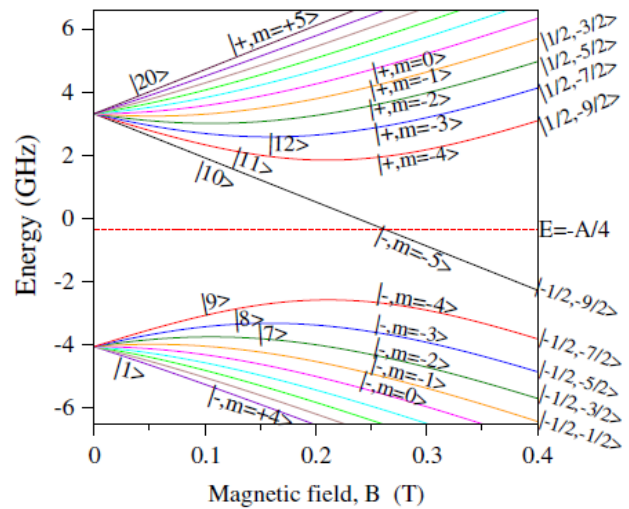

*Figure X.9. The 20 spin energy levels of Si:Bi may be labeled in alternative ways[66]: (i) in order of increasing energy |1>, |2> … |20>; (ii) by using the adiabatic basis | ±,m> of doublets as described in reference [66]; (iii) by their asymptotic, high-field form |$m_s$,$m_I$> where $m_s$ and $m_I$ are the electron and nuclear spin states respectively. States |10> and |20> are not mixed[66]. Reprinted figure with permission from M. H. Mohammady, G. W. Morley and T. S. Monteiro, Physical Review Letters, **105**, 067602 (2010). Copyright 2010 by the American Physical Society.*

Ion implantation of Bi into silicon has been demonstrated with ~100% of the implanted Bi atoms being substitutionally incorporated into the silicon lattice[77]. Scanning tunneling microscopy (STM) has been used to study the electronic states of single Bi atoms in silicon[78]. It has been proposed that STM with inelastic electron tunneling spectroscopy could allow single nuclear spin readout of $^{209}$Bi in silicon[79].

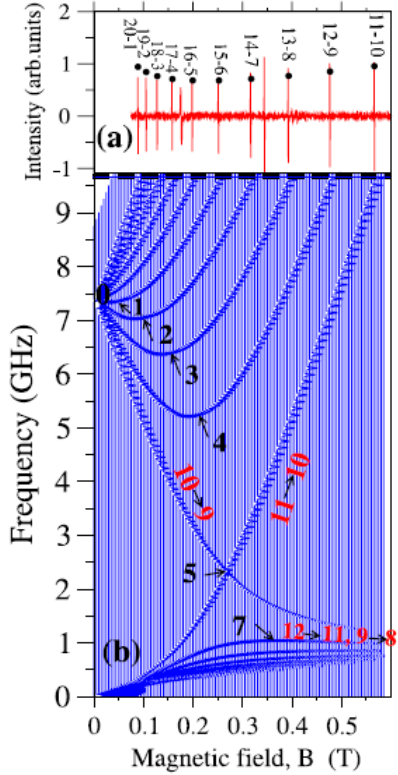

*Figure X.10. a) Comparison between theory (black dots) and experimental continuous wave EPR signal at 9.7 GHz[66]. b) Calculated EPR spectra are seen to line up with the experimental spectra at 9.7 GHz. Cancellation resonances are labelled 0,1,2,3,4,5,7 with arrows. Some transitions are labelled, e.g. "10→9" corresponds to the transition from state |10> to state |9> as defined above in Figure X.9. Reprinted figure with permission from M. H. Mohammady, G. W. Morley and T. S. Monteiro, Physical Review Letters, **105**, 067602 (2010). Copyright 2010 by the American Physical Society.*

## X.6 Electrically-Detected Magnetic Resonance (EDMR)

Electrically-detected magnetic resonance has been used with Si:P to sensitively probe nuclear spins with pulsed ENDOR at high[20, 22] and low magnetic fields[80-81]. High magnetic fields have been used to polarize Si:P electron spins[82] and this has been transferred to nuclear spins with optical excitation[18] and entirely electrically[19].

As shown in Figure X.1, EDMR has been used to show that strain is useful for tuning Si:P resonance frequencies[30].

At low magnetic fields, the mechanism for Si:P EDMR makes use of a dangling bond defect coupled to the donor, clearly demonstrated with electrically-detected pulsed EPR using two excitation frequencies[83] (electron double resonance or ELDOR). The dangling bond defects can be better understood with electrically-detected pulsed ESEEM (electron spin echo envelope modulation), and these defects limit the donor electron spin relaxation times to microseconds[84-86]. This limitation is not present at high magnetic fields as the Si:P EDMR signal is due to spin-dependent trapping of conduction electrons for which dangling bonds are not involved[82].

EDMR has allowed several different unusual experiments. For example, neutral arsenic dopants interacting with a 2D electron gas have been studied with continuous-wave EDMR at 9.7 GHz and 94 GHz[87]. The Anderson-Mott transition between conduction by sequential tunneling through isolated dopant atoms, and conduction through thermally activated impurity Hubbard bands has been studied in arrays of a few arsenic dopant atoms in a silicon transistor[88]. Single erbium spins with resolved hyperfine structure have been electrically detected after resonant optical excitation[89]. The use of the valley degree of freedom has been considered with dopants in silicon both experimentally[90-92] and theoretically[93-94]. The quantum confinement due to silicon nanowires may increase the temperatures where silicon donor quantum devices can operate[95].

## X.7 Conclusions and Outlook

Despite being currently less advanced than some other quantum technologies, donors in silicon have great potential because their mature materials science provides long coherence times and the semiconductor industry has developed techniques for wafer-scale fabrication. Atomic scale fabrication using scanning tunnelling microscopy could allow the creation of vast arrays of donors with single-shot readout using single electron transistors. Hybridizing donors in silicon with other qubits will provide advantages such as the use of donors as a quantum memory for superconducting qubits[36,96]. As with nitrogen-vacancy centres in diamond[97-98], the first useful applications of donor qubits in silicon may be as sensors rather than in quantum computation.

The best way to couple up many donor qubits remains an open question, the answer to which could set the future direction for the field.

Acknowledgements: Dane R. McCamey provided useful feedback on a draft of this chapter.

# Chapter References